\documentclass[12pt]{revtex4}

\usepackage{amsmath,graphicx,latexsym}
\usepackage{bm}
\usepackage{epsfig}
\usepackage{amsmath}
\usepackage[colorlinks,linkcolor=blue,citecolor=red]{hyperref}
\usepackage{epsfig}
\usepackage{latexsym,amsmath,amssymb}
\usepackage{graphicx}
\usepackage{slashed}
\usepackage{bm}
\usepackage{epsfig}
\usepackage{dcolumn}

\usepackage{array}
\usepackage{booktabs}
\usepackage{tabu}
\def\beq{\begin{equation}}
\def\eeq{\end{equation}}
\def\ber{\begin{eqnarray}}
\def\eer{\end{eqnarray}}
\def\benu{\begin{enumerate}}
\def\eenu{\end{enumerate}}

\def \lleq {\lower0.9ex\hbox{ $\buildrel < \over \sim$} ~}
\def \ggeq {\lower0.9ex\hbox{ $\buildrel > \over \sim$} ~}

\begin{document}

\title{Non-canonical scalar field in low anisotropy universe with intermediate inflation}

\author{Mahdeyeh Naderi}
\email{m.naderi78@yahoo.com}
\affiliation{Sanandaj Branch, Islamic Azad university, Sanandaj, Iran}
\author{Ali Aghamohammadi}
\email{a.aqamohamadi@gmail.com;a.aghamohamadi@iausdj.ac.ir}
\affiliation{Sanandaj Branch, Islamic Azad university, Sanandaj, Iran}
\author{Abdollah Refaei}
\email{abr412@gmail.com;A.Refaei@iausdj.ac.ir}
\affiliation{Sanandaj Branch, Islamic Azad university, Sanandaj, Iran}
\author{Haidar Sheikhahmadi}
\email{h.sh.ahmadi@gmail.com;h.sheikhahmadi@ipm.ir}
\affiliation{School of Astronomy, Institute for Research in
Fundamental Sciences (IPM), P. O. Box 19395-5531, Tehran, Iran}
\affiliation{Center for Space Research, North-West University, Mafikeng, South Africa}

\begin{abstract}
 The behaviour of a non-canonical scalar field within an anisotropic Bianchi type I,  spatially homogeneous, Universe in the frame work of the intermediate inflation will be studied. It will be examined on the condition that both the anisotropy and non-canonical sources come together and is there any improvement in compatibility with the observational data originated from plank $2015$?.  Based on this investigation it can be observed that automatically a steep potential which can manage inflation in a better way will be obtained. Additionally, as a common procedure for an inflationary study, we shall try to calculate the related inflationary observables such as the amplitude of the scalar perturbations, scalar and tensor spectral indices,  tensor-to-scalar ratio, the running spectral index, and the number of  e-folds. As an exciting part of our results, we will find that our model has a good consistency compared to data risen by CMB and different Planck results. To justify our claims, the well known canonical inflationary scenario in an anisotropic Bianchi type I Universe also will be evaluated.
\textit{keywords:} Low anisotropy; Non-canonical scalar field; Intermediate inflation; Steep potential
\end{abstract}

\maketitle

%\bigskip

\section{Introduction}\label{secintro}
{Based on over three decades  of large-scale investigations \cite{Sta80,Gut81,Lin82,Alb,Lin83,Lin86a,Lin86b}, obviously the inflationary paradigm  is going to be the corner stones of modern cosmology. The inflation theory can describe the early  Universe evolutions successfully and also could be considered as a remedy for three vital problems which old big bang theory was faced i.e.   the flatness,  horizon and  heavy monopoles problems \cite{Lid00,Bas,Lem,Kin,Bau09,Bau14}.
Besides, it seems to obtain a correct ratio  for tensor-to-scalar ratio and, in general, a correct behavior of primordial perturbations this model   is requisite \cite{Liddle0,Langl,Lyth,Guth00,Lidsey97,Bas,Mukhanov-etal,Haidar,Haidar2}.
 In the standard inflation model, the potential term of Lagrangian is dominated comparing to the canonical kinetic term, i.e. the potential term dominated during inflation \cite{Lid00,Bas,Bau09,Bau14}. However there exist inflationary models in which the kinetic term has different form  from  the canonical one, namely non-canonical models, such as Dirac-Born-Infeld (DBI) action  where the non-canonical kinetic term is attributed to the scalar field. It could be realized  that DBI scalar field model can be assumed as a subset of k-inflation scenario \cite{Arm,Gar,Li,Hwa,Fra10a,Fra10b,Unn12,Unn13,Zha14a,Gol,Nazavari}. The observational constraints on  $k$-inflation and its perturbations have been considered in literature \cite{Gar} and \cite{Li}.
 Here we should emphasis despite a huge number of inflationary models, the precise data originated from inspecting Cosmic Microwave Background, CMB, have been reduced dramatically the number of allowable  inflationary models \cite{Mar13,Mar14,Hos14a}.
Additionally, some other noticeable researches have been worked out in the context of non-canonical inflationary scenario and we refer the reader to \cite{Hwa,Fra10a,Fra10b,Unn12,Unn13,Zha14a,Gol}.
In \cite{Unn12} it was shown that one can reduce the values of slow-roll parameters  and accordingly the slow-roll regime condition can be obtained very easily by using of a non-canonical Lagrangian than the canonical case. In addition, it has been shown that the steep potentials connected to dark energy in the canonical setting can drive inflation in the non-canonical framework precisely \cite{Unn12}. In the  non-canonical setup, the power law inflation is consistent with the observational results and one can obtain a way to end the inflation and it causes to get rid of changing the form of the power law of the scale factor surrounding  the horizon exit \cite{Unn13}.\\
Mostly there are exist various ways  to, in the slow-rolling inflationary scenarios,  obtain expression of observables such as the tensor and scalar spectral indices, the running of them and tensor-to-scalar ratio. Amongst them one can refer to the introducing different types for scale factor to run inflation and then examining the results comparing to observations. Here we are interesting to the intermediate type, which is the most generic and well-known, to investigate its role to derive inflation  and for the first  time has been presented by \cite{Barrow11}.
In this procedure, the scale factor introduced as an exponential function based on the cosmic time, i.e.  $a(t) = \exp\big( \kappa t^f \big)$, $\kappa>0$ where usually $0<f<1$ \cite{Vallinotto,Starobinsky}. It leads to  an asymptotically negative power-law potential, we can refer the steep potentials for instance \cite{Rendall}. Here we can supply the reason why people named these set of scale factors the intermediate ones.  For them usually the expansion of the Universe is faster than the case which made by power-law one, i.e. ($a(t)=t^p, p>1$), and slower than de-Sitter inflation ($a(t)=\exp(Ht), H=constant$)}.  It is interesting also we mention here that in Einstein  gravity intermediate inflation  for $\alpha=2/3$  creates scale invariant perturbations
 \cite{Barrow11,Barrow-etal,Vallinotto,Starobinsky}. One important reason to consider the intermediate inflation is  its results for tensor-to-scalar ratio and scalar spectrum index which are in a good agreement comparing to the CMB data \cite{BarrowNunes}. Due to advantages of the intermediate proposal  in solving  problems of inflation this scenario preserve an appropriate place in the community and for more details we can refer the reader to the literature \cite{Muslimov,BarowLiddle02,BarrowLiddle,kk,mohammadi} and references there.\\
The  majority of investigations  to find out  dynamical evolution of inflationary models have been  done in a homogenous and isotropic background for instance Friedmann-Lemaitre-Robertson-Walker, FLRW, metric. However, a bit little deviation of isotropy at the level of $10^{-5}$, has also been proposed  by Bennett \textit{et al}, and subsequently this suggestion was  approved by high resolution Wilkinson Microwave Anisotropy Probe, WMAP \cite{ch1:9,WM2}. We should emphasise here although according to recent  studies the anisotropy should be small its imprints in large scale structure formation are considerable. To justify this claim the the effects of anisotropy on the early Universe evolutions and especially the primary seeds of structure formation in the frame work of  Bianchi type I (BI) exactly have been investigated  \cite{ko,ku,Y1,17,18,aghaohamadi}.  Amongst  the Bianchi  different types we can refer to the Kasner-type as a specific one; in which cosmological scale factors evolve by a power-law function of time. In General Relativity, GR, the vacuum Kasner solutions \cite{r7} and their fluid filled counterparts, the
BI models, were verified effectively as a starting point for the investigation of the
structure of anisotropic models. Barrow and Clifton  \cite{r8,r9} have shown  that it
 is  possible finding out the solutions of the Kasner type for $R^n$-gravity.
Newly, the authors in \cite{18a} have discussed the effects of low anisotropy on the interacting Dark Energy, DE, models and  have shown the advantages of their model comparing to the  standard FLRW, $\Lambda$ Cold Dark Matter, $\Lambda$CDM, and $w$CDM model results. Additionally, they showed  that the anisotropy should get a non-zero value at the present time. Let's again turn our attention to the BI Universe; in fact the BI model is a straightforward extension of the flat FLRW metric so we can emphasize that it is the  simplest model of anisotropic but homogenous Universe with spatial flatness. Against the FLRW Universe, which has a same scale factor for its three spatial directions, in the BI Universe the scale factor could change in different independent directions. Hence the study of inflation in an anisotropic Universe has much more advantages than isotropic one. Therefore, based on aforementioned reasons, in this work we are going to consider the  anisotropic model to investigate the effects of the intermediate inflation with a non-canonical scalar field  \cite{BarowLiddle02,BarrowLiddle}.
By the way, here we want to answer this question why we need to deal with non-canonical model instead of pure canonical one. An immediately and temporary answer may be is  intermediate inflation in the standard canonical inflation  has some drawbacks and faced to failure comparing with observations. In Refs.\cite{BarowLiddle02,BarrowLiddle} it was showed that in light of the observations risen from  the Cosmic Background Explorer ,COBE,  the scalar and tensor
power spectral  expressed by  intermediate inflation has no any valuable results. Besides, the aforementioned drawbacks  without adding some extra processes to the model the intermediate inflation never cease, behaves as same as the eternal inflation \cite{BarrowLiddle}.
At present work, we want to seek a probable remedy for these problems in which
the canonical, more even non-canonical, inflation with  intermediate inflation faced but in an anisotropic framework.\\
This work is organized as follows: In Sec.\,\ref{secnon-can}, we will express the main dynamical equations for non-canonical Lagrangian in an anisotropic metric. And in Sec.\,\ref{secInfnon-can} by virtue of an intermediate scale factor we will evaluate  the inflationary observables and will compare their results with the  Planck $2015$ data as a well-known criterion. Also, for the asymptotical regimes, i.e.  canonical intermediate inflation and isotropic background, we will show that their results are not in a good agreement compared to general at hand proposal and the planck data. At last, Sec.\,\ref{seccon} is devoted to conclusion and discussions.

\section{ non-canonical model in an anisotropic metric}\label{secnon-can}
Usually non-canonical inflation could be expressed by the following action
\begin{equation}
\label{action}
S = \int {{{\rm{d}}^4}} x\;\sqrt{-g}~\mathcal{L}(X,\phi),
\end{equation}
where  Lagrangian ${\cal L}$ is a function of scalar field $\phi$ and its derivatives, i.e. the kinetic  term $X \equiv {g^{\mu\nu}\nabla _\mu
}\phi {\nabla_\nu }\phi /2$.  By varying the action and after some algebra
the equations of energy density
$\rho_{\phi}$ and pressure $p_{\phi}$ are obtained as follows: \cite{Arm,Gar,Li,Hwa,Fra10a,Fra10b,Unn12,Unn13,Zha14a}
\begin{eqnarray}
\label{rhodef}
{\rho _\phi } &=&
2X\left( {\frac{{\partial {\cal L}}}{{\partial X}}} \right) - {\cal
L}~,
\\
\label{pdef}
{p_\phi } &=& {\cal L}~.
\end{eqnarray}
\label{sec2}
As  mentioned in introduction the BI cosmology refers to a spatially homogeneous background but not necessarily isotropic one. As a remembrance please note that we will consider BI cosmology in entire of this work expect when it has mentioned obviously. The metric of the BI model could be given by
\begin{equation}\label{1}
ds^2=dt^{2}-A^{2}(t)dx^{2}-B^{2}(t)dy^{2}-C^{2}(t)dz^{2},
\end{equation}
where the metric components $A, B$ and $C$ are merely  functions of time, for more details about Lie algebra and isometry group of the BI metric we refer the reader to  \cite{35a} . From the literature we know the energy momentum tensor for perfect fluid is expressed by
\begin{eqnarray}\label{4}
T^{\mu}_{\nu}=diag[\rho,-p,-p,-p],
\end{eqnarray}
where $\rho$ and $p$ represent the energy density and pressure respectively.  Additionally, the field equations in the axial symmetry BI metric are obtained as \cite{36,37,38}
\begin{eqnarray}
3H^{2}-\sigma^{2}&=&\frac{1}{M_p^2}(\rho_{\phi}),   \label{Fri} \\
3H^2+2\dot{H}+\sigma^{2}&=&-\frac{1}{M_p^2}\left(p_{\phi}\right), \label{Fri2}
\end{eqnarray}
where $M_p^2=1/(8\pi G)$ is the reduced Planck mass, and $\sigma^2=\sigma_{ij}\sigma^{ij}/2 $  in which
$\sigma_{ij}=u_{i,j}+\frac{1}{2}(u_{i;k}u^{k}u_{j}+u_{j;k}u^{k}u_{i})+\frac{1}{3}\theta(g_{ij}+u_{i}u_{j})$
is the shear tensor.  By virtue of this tensor we can write down $(\sigma_{ij}u^j=0, \sigma^i_{~i}=0)$ that  describes the rate of distortion of the
matter flow. The  scalar expansion introduced by $3H=u^{j}_{;j}$  where $u^j$ is 4-velocity and in a comoving coordinate it is given by   $(u^i=\delta^i_0)$.
Also the components of the  Hubble parameter and the shear tensor based on the  Eq.(\ref{1}) are expressed as \cite{36}
\begin{eqnarray}
H&=&\frac{1}{3}(\frac{\dot{A}}{A}+\frac{\dot{B}}{B}+\frac{\dot{C}}{C}),   \label{14} \\
\sigma^{2}&=&3H^2-(\frac{\dot{A}\dot{B}}{AB}+\frac{\dot{B}\dot{C}}{BC}+\frac{\dot{A}\dot{C}}{AC}). \label{15}
\end{eqnarray}
If one takes $A=B^{\lambda}$ with $B=C$ the scale factor is appeared as $a = (ABC)^{1/3}=(B)^{(\lambda+2)/3} $ where  $\lambda$ is a real constant. Then by  assuming $H_2 = \frac{\dot{B} }{B}$ the Hubble parameter and the shear are reduced to the following simplified equations
\begin{eqnarray}
H = \frac{2 + \lambda }{3}H_2,\label{1an}\\
\sigma^2 = \frac{(\lambda-1)^2H_2^2}{3}.\label{2an}
\end{eqnarray}
By combining the Friedmann equations and embedded them into the Klein-Gordon equation the  conservation equation is resulted as
\begin{equation}
\label{rhophidot}
{\dot \rho _\phi } + 3H\left( {{\rho _\phi } + {p_\phi }} \right) = 0.
\end{equation}
\section{ Intermediate inflation for an anisotropic Universe  and non-canonical Lagrangian}\label{secInfnon-can}
\begin{figure}[ht]
\centering
\includegraphics[scale=.50]{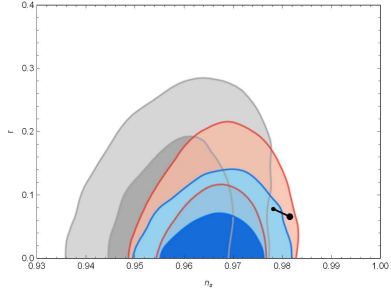}
\caption{{\it{The $r-n_s$ diagram indicates  prediction of
the non-canonical intermediate inflationary model in anisotropy background for the specified values of  $\alpha=3$, $\kappa=3.02\times 10^{-12}$,  $f=10^{-4}$ ,$M_p=10^{18}$, $M=10^{12}$ and $\mu=10^{-6}$,
in comparison with the observational results of Planck 2015. Where the likelihood  of  Planck 2013 (grey contours), Planck TT+lowP(red contours), Planck TT,TE,EE+lowP(blue contours) and   the thick black line
indicate the predictions of our model in which small and large dots are the value of $n_s$ at the number of e-fold $N=55,~N=66$.}}}
\label{fignsr}
\end{figure}

In this section the inflationary behaviour for an intermediate scale factor by means of extended canonical Lagrangian inside a BI Universe is studied. Now let us turn our attention to investigate the inflationary evolution in the aforementioned framework. To do so we want to begin with introducing  the namely first and second slow-roll parameters, viz.
\begin{eqnarray}
\label{eps}
\varepsilon  &=&  - \frac{{\dot H}}{{{H^2}}},
\\
\label{eta}
\eta  &=& \frac{\dot{\varepsilon}}{H \varepsilon}.
\end{eqnarray}
To receive an accelerating phase as a necessary part of initial Universe, i.e. $\ddot a>0$,  from Eq.(\ref{eps}) one immediately realize that the first slow-roll parameter should behave like  $\varepsilon<1$. Also as mentioned above one of the big triumphes of inflation paradigm was the finding a remedy to cope with the horizon problem; inflation should drag on in order to  persist the relation of $\varepsilon<1$  but the acceleration gets much smaller amounts than unity to run inflation. Hence, inflation occurs and persists if and only if  both $\varepsilon $ and
$\left| \eta  \right|$ being much less than unity and so  these assumptions in the literature named usually the slow-roll approximations. Another critical parameter to drive inflation in expected way is the number of e-fold which is defined as
\begin{equation}
\label{N}
N  = \int_{t_i}^t H dt = \int_{\phi_i}^{\phi} \frac{H}{{\dot \phi }}d\phi.
\end{equation}
In order to solve the horizon problem  the number of e-fold should at least  become  more than 60 \cite{kk}. Now after introducing the necessary instruments of running the inflation we can go back to the Lagrangian again. The Lagrangian density  which we shall consider can be considered as  the following \cite{Unn12,Unn13}
\begin{equation}
\label{Lag}
\mathcal{L}(X,\phi ) = X{\left( {\frac{X}{{{M^4}}}}
\right)^{\alpha  - 1}} - \;V(\phi ),
\end{equation}
where  $M$ has the  dimension of mass and  $\alpha$ is a  dimensionless parameter introduced to  afford turning about to canonical case, i.e. ${\cal
L}(X,\phi ) = X - V(\phi )$. Additionally, the  Lagrangian \ref{Lag} satisfies the requirements
$\partial {\cal L}/\partial X \ge 0 $ and ${\partial ^2}{\cal
L}/\partial {X^2} > 0$   to cope with both the null-energy condition and  physical propagations of perturbations respectively \cite{Fra10a}. This type of Lagrangian has been taken in account in vast number of prior literature
 to  investigate some   steep
potentials for chaotic or other inflationary scenarios \cite{Unn12}. To refine the power law
inflation in light of Planck $2013$ this Lagrangian has been considered as well \cite{Unn13}.\\
Now let's start the calculations based on the  Lagrangian introduced in Eq.(\ref{Lag}). To do so we want to substitute  the Lagrangian (\ref{Lag}) into the Eqs. (\ref{rhodef}) and (\ref{pdef}) and thence the energy density and pressure of the scalar field $\phi$ are given by
\begin{eqnarray}
\label{rhophi}
{\rho _\phi } &=& \left( {2\alpha - 1}
\right)X{\left( {\frac{X}{{{M^4}}}} \right)^{\alpha  - 1}} +
V(\phi),
\\
\label{p}
{p_\phi } &=& X{\left( {\frac{X}{{{M^4}}}} \right)^{\alpha  - 1}} - V(\phi ).
\end{eqnarray}
In addition the dynamical equation of the scalar field, i.e. Klein-Gordon equation, by embedding  the Eqs.(\ref{rhophi}) and (\ref{p}) into the conservation equation
(\ref{rhophidot}) will be expressed as follows
\begin{equation}
\label{phiddot}
\ddot \phi  + \frac{{3H\dot \phi }}{{2\alpha  - 1}}
+ \left( {\frac{{V'(\phi )}}{{\alpha (2\alpha  - 1)}}}
\right){\left( {\frac{{2{M^4}}}{{{{\dot \phi }^2}}}} \right)^{\alpha
- 1}} = 0.
\end{equation}
Whereas we have no any interaction of type of non-minimally coupled chameleonic mechanism here, so by varying the lagrangian with respect to the scalar field we can obtain the above equation equally to the procedure which have been  used in the papers \cite{Kho2,MOF1,ch19,ch20,ch1:20a}.  Now by substituting  Eqs. (\ref{14},\ref{15},\ref{1an},\ref{2an})  into the Eqs.(\ref{rhophi}) and (\ref{p}), the slow-roll parameters, i.e. Eqs.(\ref{eps}) and (\ref{eta}),   based on the potential $V(\phi)$  are expressed as follows
\begin{eqnarray}
\label{epsV}
{\varepsilon _V} &=&\frac{\sqrt{3(2\lambda +1)}}{2+\lambda} {\left[ {\frac{1}{\alpha }{{\left(
{\frac{{3{M^4}}}{V(\phi)}} \right)}^{\alpha  - 1}}{{\left(
{\frac{{{M_P}V'(\phi)}}{{\sqrt 2 \;V(\phi)}}} \right)}^{2\alpha }}}
\right]^{\frac{1}{{2\alpha  - 1}}}},
\\
\label{etaV}
{\eta _V} &=& \frac{\sqrt{3(2\lambda +1)}}{2+\lambda}\left( {\frac{{\alpha {\varepsilon
_V}}}{{2\alpha  - 1}}} \right)\left( {\frac{{2V(\phi )V''(\phi
)}}{{V'{{(\phi )}^2}}} - 1} \right),
\end{eqnarray}
The Eqs.(\ref{epsV}) and (\ref{etaV}), so called the first and second potential based
slow-roll parameters respectively. In addition,  the  slow-roll  approximation implies
the potential energy should be more larger than the kinetic one and therefore the
Friedmann equation (\ref{Fri}) is reduced to
\begin{equation}
\label{Frisr}
H^2\left(\phi\right) =\frac{(2+\lambda)^2}{9(2\lambda+1)}
\frac{1}{{M_P^2}}V(\phi ).
\end{equation}
Meanwhile, under the slow-roll condition the dynamical equation of the
scalar field, (\ref{phiddot}), is took the form
\begin{equation}
\label{phidot}
\dot \phi  =  -
\theta {\left \{\frac{\sqrt{3(2\lambda+1)}}{2+\lambda} {\left( {\frac{{{M_P}}}{{\sqrt 3 \alpha }}}
\right)\left( {\frac{{\theta V'(\phi )}}{{\sqrt {V(\phi )} }}}
\right){{\left( {2{M^4}} \right)}^{\alpha  - 1}}}
\right\}^{\frac{1}{{2\alpha  - 1}}}},
\end{equation}
where $\theta  = \pm 1$ when the sign of $V'(\phi ) $ is $\pm$ \cite{Unn12, kk}.
As mentioned in the aforementioned sections, the main aim of this study goes back to investigate the intermediate
inflation in an anisotropic Universe, i.e. BI Universe. The scale factor expressed as $a = (ABC)^{1/3}=(B)^{(\lambda+2)/3} $ in which parameter $\lambda$  introduced to indicate the deviations of the isotropic background and could be considered a little bit larger or smaller than unity. Hence the appellation of  low anisotropy implies these small deviations; and the component  $B$ of the metric in intermediate inflation  is expressed as
\begin{equation}
\label{at}
B(t) = a_i \exp \left[ {{\kappa }{{\left( {{M_P}t} \right)}^f}}
\right],
\end{equation}
where $a_i$ is the scale factor  in $y$ axis direction, i.e. the $g_{22}$ component of the metric tensor  at the initial time of the inflation. Thereupon, one will be able to obtain the main scale factor as
$a=(a_i \exp [ {{\kappa ^2}{{\left( {{M_P}t} \right)}^f}}
])^{(\lambda+2)/3}$.
 Signally by virtue of this definition, the parameters of Hubble and shear could be obtained  as follows
\begin{equation}
\label{Hubble}
H^2=\frac{{{\kappa}^2{f^2}{{({M_p}t)}^{2f}}{{(2 + \lambda )}^2}}}{{9{t^2}}},
\end{equation}
and
\begin{equation}
\label{shear}
\sigma^2=\frac{{{\kappa}^2{f^2}{{({M_p}t)}^{2f}}{{( - 1 + \lambda )}^2}}}{{3{t^2}}}.
\end{equation}
In the above expressions we have the constraints ${\kappa }>0$ and $0<f<1$ \cite{BarowLiddle02,BarrowLiddle}. For more convenient the scale factor is normalized to  the present time values   as $a_0=1$. By using the Eqs.(\ref{Fri}) and (\ref{Fri2}) with the
intermediate scale factor (\ref{at})  one receives
\begin{eqnarray}
\label{rhot}
\rho _\phi &=& \frac{{\kappa }^{2f}M_p^2(M_pt)^{2f}(1 + 2\lambda )}{t^2}
,
\\
\label{pt}
{p_\phi } &=&  - \frac{{{\kappa}fM_p^2{{({M_p}t)}^f}[2( - 1 + f)(2 + \lambda ) + {\kappa }f{{({M_p}t)}^f}(5 + 2\lambda (1 + \lambda ))]}}{{3{t^2}}}.
\end{eqnarray}
Considering the slow-roll condition, i.e. ${\rho _\phi } = V(\phi )$, and  Eq.(\ref{rhot}) we obtain
\begin{equation}
\label{Vt1}
V(\phi )=\frac{{{\kappa ^2}{f^2}{M_P^2}{{\left( {M_P t} \right)}^{2f}}\left( {1 + 2\lambda } \right)}}{{{t^2}}}.
\end{equation}
Substituting  Eq.(\ref{Vt1}) into (\ref{phidot}) we receive  a first order differential equation to the scalar field  as  follows
\begin{equation}
\label{phit}
\dot \phi(t)  = {\left( { - \frac{{{2^\alpha }\left( { - 1 + f} \right){{\left( {{M^4}} \right)}^{ - 1 + \alpha }}{M_P}\sqrt {1 + 2\lambda } \sqrt {\frac{{{\kappa ^2}{f^2}M_P^2{{\left( {{M_P}t} \right)}^{2f}}\left( {1 + 2\lambda } \right)}}{{{t^2}}}} }}{{\alpha \left( {2 + \lambda } \right)t}}} \right)^{\frac{1}{2}/\alpha }}.
\end{equation}
Now, by integrating  Eq.(\ref{phit}) and after some manipulations time $t$ could be obtained  as a function of $\phi$,
\begin{eqnarray}
\label{tphi} \nonumber
t(\phi ) &= &{2^{ - \frac{{2\alpha }}{{ - 2 + f + 2\alpha }}}}\\
&\times&{\left( { - \frac{{\left( {2 - f - 2\alpha } \right)\phi }}{{{\left( { - \frac{{{2^\alpha }\left( { - 1 + f} \right){{\left( {{M^4}} \right)}^{ - 1 + \alpha }}{M_P}\sqrt {1 + 2\lambda } \sqrt {{\kappa ^2}{f^2}M_P^{2 + 2f}\left( {1 + 2\lambda } \right)} }}{{\alpha \left( {2 + \lambda } \right)}}} \right)}^{  \frac{1}{2}/\alpha }}\alpha }} \right)^{\frac{{2\alpha }}{{ - 2 + f + 2\alpha }}}}.
\end{eqnarray}
Then to find the form of the potential, we can substitute the above solution in Eq.(\ref{Vt1}) and it gives
\begin{equation}
\label{Vt2}
V(\phi ) = {V_0}{\phi ^s},
\end{equation}
where
\[\begin{array}{l}
{V_0} = {\kappa ^2}{f^2}M_P^4\left( {1 + 2\lambda } \right)\\
 \times \left( {{M_P}{{\left( { \frac{{-\left( {2 - f - 2\alpha } \right)}}{{2\alpha {{\left( {  \frac{{{2^\alpha }\left( {  1 - f} \right){{\left( {{M^4}} \right)}^{ - 1 + \alpha }}\sqrt {{\kappa ^2}{f^2}M_P^{4 + 2f}{{\left( {1 + 2\lambda } \right)}^2}} }}{{\alpha \left( {2 + \lambda } \right)}}} \right)}^{\frac{1}{2}/\alpha }}}}} \right)}^{\frac{{2\alpha }}{{ - 2 + f + 2\alpha }}}}} \right)^{-2+2f}\equiv { V_0}^\ast{\kappa ^\frac{4\alpha-2}{-2+2\alpha+f}}
\end{array}\]
is a constant and
\begin{equation}
\label{s}
s = \frac{{2\alpha ( - 2 + 2f)}}{{ - 2 + f + 2\alpha }}.
\end{equation}
It is obvious that the achieved potential behaves like the  power law potentials \cite{BarowLiddle02,BarrowLiddle}. Whereas  the value of the parameter $f$  for the intermediate scale factor (\ref{at}) gets the values of betwixt $0$ and $1$ \cite{Muslimov,BarowLiddle02,BarrowLiddle,kk,mohammadi}. Thence, from Eq.(\ref{s}) one can conclude that  the parameter $s$ in (\ref{Vt2})  must be in the range $0 < s <-2\alpha/(\alpha - 1)$  to authorize the existence of  intermediate inflation which  $\alpha>1$ according to equation (\ref{Lag}). Since  in the standard canonical setting ($\alpha =1$), so the $s$ parameter is varying between   $-\infty< s <0$.
Now, given the  inverse power law potential form  as source of inflation  in the slow roll condition,
 we can obtain the necessary relations
  for determining  the inflationary observables.
The expression of  the scalar and tensor power spectrum in the slow roll  regime  are given  as \cite{Gar}
\begin{eqnarray}
\label{psk}
{{\cal P}_s}&=&(\frac{H^2}{2\pi(c_s(\rho_{\phi}+p_{}\phi)^{1/2}})^2_{aH_{iso}=c_sk}
,
\\
\label{ptk}
{{\cal P}_t} &=&\frac{8}{M_p^2}(\frac{H}{2\pi})^2_{aH_{iso}=k}.
\end{eqnarray}
By considering the Lagrangian (\ref{Lag}) and also the Eqs.(\ref{1an}, \ref{epsV}) in anisotropic metric, the above equations are expressed as
\cite{Unn12,Unn13}
\begin{eqnarray}
\label{Ps}
{{\cal P}_s} &=&\frac{(2+\lambda)^3}{(3(2\lambda+1))^3/2} \frac{1}{{72{\pi ^2}{c_s}}}\left(
{\frac{{{6^\alpha }\alpha V{{(\phi )}^{5\alpha  -
2}}}}{{M_P^{14\alpha  - 8}{{ M}^{4(\alpha  - 1)}}V'{{(\phi
)}^{2\alpha }}}}} \right)_{a{ani} = {c_s}k}^{\frac{1}{{2\alpha  - 1}}}.\\
{{\cal P}_t}&=&\frac{(2+\lambda)^2}{3(2\lambda+1)}\Big(\frac{2V(\phi)}{3{\pi}^2 M_p^4}\Big)_{aH_{ani}= k}.\label{PTV}
\end{eqnarray}
To receive equations \ref{Ps} and \ref{PTV}  we used  $H_{ani}=\frac{2+\lambda}{\sqrt{3(2\lambda+1)}}H_{iso}$ where subscribes ${ani} $ and ${iso}$ refer to the anisotropic  and isotropic respectively. Now let's explain a little bit more about the constraint $aH_{ani} = {c_s}k$ in above equations. In fact, based on leading literature and textbooks the scalar power spectrum should be assessed  at the sound horizon exit that specified by $aH_{ani} = {c_s}k$ where $k$ is the comoving wave number and
$c_{s}$ refers to the sound speed \cite{Arm,Gar,Li,Hwa,Fra10a,Fra10b,Unn12,Unn13,Zha14a,kk}. Additionally, the sound speed has a definition as follows
\begin{equation}
\label{csdef} c_s^2 \equiv \frac{{\partial {p_\phi }/\partial
X}}{{\partial {\rho _\phi }/\partial X}} = \frac{{\partial {\cal
L}(X,\phi )/\partial X}}{{\left( {2X} \right){\partial ^2}{\cal
L}(X,\phi )/\partial {X^2} + \partial {\cal L}(X,\phi )/\partial
X}}.
\end{equation}
And for our investigation here it takes the following form
\begin{equation}
\label{cs}
{c_s} = \frac{1}{{\sqrt{2\alpha  - 1} }},
\end{equation}
where behaves just as a constant. Replacing  the potential (\ref{Vt2}) onto Eqs.(\ref{Ps}) and (\ref{PTV})
after some algebra gives
\begin{eqnarray}\label{Psphi}
{P_s} &= &{(\frac{{\left( {2 + \lambda } \right)}}{{\sqrt {3\left( {2\lambda  + 1} \right)} }})^3}\\ \nonumber
 &\times &\frac{{{{\left( {{6^\alpha }M_P^{8 - 14\alpha }{\mkern 1mu} \alpha {\mkern 1mu} {\mu ^{4 - 4\alpha }}{{\left( {\frac{{2\alpha \left( { - 2 + 2f} \right)}}{{ - 2 + f + 2\alpha }}} \right)}^{ - 2\alpha }}{{\left( {{V_0}} \right)}^{ - 2 + 3\alpha }}} \right)}^{\frac{1}{{ - 1 + 2\alpha }}}}}}{{72{\pi ^2}{c_s}}}(\phi )_{aH_{ani} = {c_s}k}^{\frac{{\alpha \left( {6f - 4} \right)}}{{2\alpha  + f - 2}}},
\end{eqnarray}
and
\begin{equation}
\label{PTphi}
{{\cal P}_t} = {(\frac{{\left( {2 + \lambda } \right)}}{{\sqrt {3\left( {2\lambda  + 1} \right)} }})^2}\frac{{2{V_0}}}{{3M_P^4{\pi ^2}}}(\phi )_{aH_{ani} = k}^{\frac{{4\left( { - 1 + f} \right)\alpha }}{{ - 2 + f + 2\alpha }}},
\end{equation}
where $\mu= M/M_p$. From Eq.(\ref{Psphi}), it obviously could be seen that for the value of $f=2/3$ the
scalar power spectrum is appeared as an independent parameter of $\phi$ and so
it makes  sense like the scale-invariant Harrison-Zel$'$dovich spectrum. Now, in order to calculate the evolution of power spectrum based on  $N$, we need the scalar field in terms of the number of  e-folds. Hence, we might need to calculate the values of scalar field at the initiation of the inflation namely $\phi_{i}$. To do this end according to   the  slow-roll parameter definition i.e. equation (\ref{epsV}) we have
\begin{equation}
\label{epsilonphi}
{\varepsilon _V}=\sqrt {3\left( {2\lambda  + 1} \right)} \frac{{{{\left( {{\alpha ^{ - 1}}{2^{ - \alpha }}{3^{ - 1 + \alpha }}M_P^{2\alpha }{{\left( {\frac{{2\alpha \left( { - 2 + 2f} \right)}}{{ - 2 + f + 2\alpha }}} \right)}^{2\alpha }}{{\left( {\frac{{{M^4}}}{{{V_0}}}} \right)}^{ - 1 + \alpha }}} \right)}^{\frac{1}{{ - 1 + 2\alpha }}}}}}{{2 + \lambda }}{\phi ^{\frac{{s - \alpha s - 2\alpha }}{{ - 1 + 2\alpha }}}}.
\end{equation}
Now we  can rely on these facts that at the beginning of the inflation $\varepsilon _V=1$   therefore easily we obtain the related value of  scalar field as
\begin{equation}
\label{phibegin}
\phi_i = {{\Big[}{{\Big(}{\alpha ^{ - 1}}{2^{ - \alpha }}{3^{ - 1 + \alpha }}{M^{4\alpha  - 4}}{V_o}^{1 - \alpha }M_p^{2\alpha }{\Big)}^{ - \frac{1}{{ - 1 + 2\alpha }}}}\chi{\Big]}^{\frac{{1 - 2\alpha }}{{ - s + 2\alpha  + s\alpha }}}},
\end{equation}
where $\chi  = \frac{{(2 + \lambda )}}{{\sqrt {3(2\lambda  + 1)} }}$. By bringing in account the  Eq.(\ref{N}) we get
\begin{equation}
\label{NN}
\phi  = {{\Big(}\phi_i^{\frac{{2 - f}}{{ - 2 + f + 2\alpha } + s}} + \frac{N}{\Lambda }{\Big)}^{\frac{{ - 2 + f + 2\alpha }}{{(2 - f)(1 - s) + 2\alpha s}}}},
\end{equation}
where   $$\Lambda  = \frac{{2\alpha (2 + \lambda )\sqrt {{V_0}} {{({\gamma ^{ - \frac{{2\alpha }}{{ - 2 + f + 2\alpha }}}})}^{ - \frac{{ - 2 + f}}{{2\alpha }}}}}}{{3{M_p}( - 2 + f + 2\alpha )(1 + \frac{s}{2} - \frac{{ - 2 + f}}{{ - 2 + f + 2\alpha }})\gamma \sqrt {1 + 2\lambda } }},$$ and
$$\gamma  = \frac{{2\alpha {{( - \frac{{{2^\alpha }( - 1 + f){M^{ - 4 + 4\alpha }}{M_p}\sqrt {1 + 2\lambda } \sqrt {{\kappa ^2}{f^2}{M_p}^{2 + 2f}(1 + 2\lambda )} }}{{\alpha (2 + \lambda )}})}^{1/2\alpha }}}}{{ - 2 + f + 2\alpha }}.$$
 Now, by virtue  of Eqs.(\ref{Psphi}) and (\ref{NN}),
  the scalar power spectrum in
terms of the number of  e-folds is given by
\begin{eqnarray}
\label{Psk}
{{\cal P}_s}={(\frac{{\left( {2 + \lambda } \right)}}{{\sqrt {3\left( {2\lambda  + 1} \right)} }})^3}\frac{{{{\left( {{6^\alpha }M_P^{8 - 14\alpha }\,\alpha \,{\mu ^{4 - 4\alpha }}{{\left( {\frac{{2\alpha \left( { - 2 + 2f} \right)}}{{ - 2 + f + 2\alpha }}} \right)}^{ - 2\alpha }}{{\left( {{V_0}} \right)}^{ - 2 + 3\alpha }}} \right)}^{\frac{1}{{ - 1 + 2\alpha }}}}}}{{72{\pi ^2}{c_s}}}\times\cr
{{\Big(}{\Big(}\phi_i^{\frac{{2 - f}}{{ - 2 + f + 2\alpha } + s}} + \frac{N}{\Lambda }{\Big)}^{\frac{{ - 2 + f + 2\alpha }}{{(2 - f)(1 - s) + 2\alpha s}}}}{\Big)}^{\frac{{\alpha \left( {6f - 4} \right)}}{{2\alpha  + f - 2}}}.
\end{eqnarray}
Whereas both $H$, in the slow roll inflation, and $c_s$, in this work are constant. Consequently at the sound horizon exit, $aH = c_s k$, we have \cite{Unn13}
\begin{equation}
\label{exithorizon}
\frac{{\rm{d}}}{{{\rm{d ln k}}}} \simeq -\frac{{\rm{d}}}{{{\rm{dN}}}}.
\end{equation}
The importance of this relation goes back to calculation the especially spectral indices. For the scalar spectral index we can write
\begin{equation}
\label{nsdef}
{n_s} - 1 \equiv \frac{{d\ln {{\cal P}_s}}}{{d\ln k}},
\end{equation}
in which by using Eq. (\ref{Psk}) we get
\begin{eqnarray}
\label{nsk}
n_s&=&1 - \frac{{( - 4 + 6f)\alpha }}{{(2 - f)(1 - s) + 2\alpha s)\Lambda }}\cr
&\times&\frac{1}{{\frac{N}{\Lambda } + {{\Big{(}{{({{(\frac{{{2^{ - \alpha }}{3^{ - 1 + \alpha }}{M^{4( - 1 + \alpha )}}V_0^{1 - \alpha }{{({M_p}s)}^{2\alpha }}}}{\alpha })}^{\frac{1}{{1 - 2\alpha }}}}\chi )}^{\frac{{1 - 2\alpha }}{{s( - 1 + \alpha ) + 2\alpha }}}}\Big{)}}^{s + \frac{{2 - f}}{{ - 2 + f + 2\alpha }}}}}}.
\end{eqnarray}
Another important parameter one can refer,  to investigate the behavior and evolution of the initial cosmos, is the running parameter. We can consider the running of the scalar spectral index as
\begin{eqnarray}
\label{dnsk}
\alpha_s&=&\frac{{d{n_s}}}{{d\ln k}} =\frac{{( - 4 + 6f)\alpha }}{{{\Lambda ^2}(2 + f( - 1 + s) + 2s( - 1 + \alpha ))}}\cr
&\times&
{{\Bigg[\frac{N}{\Lambda } + {{\Big{(}{{({{(\frac{{{2^{ - \alpha }}{3^{ - 1 + \alpha }}{M^{4( - 1 + \alpha )}}V_0^{1 - \alpha }{{({M_p}s)}^{2\alpha }}}}{\alpha })}^{\frac{1}{{1 - 2\alpha }}}}\chi )}^{\frac{{1 - 2\alpha }}{{s( - 1 + \alpha ) + 2\alpha }}}}\Big{)}}^{s + \frac{{2 - f}}{{ - 2 + f + 2\alpha }}}}\Bigg]^{-2}}}
\end{eqnarray}
After the completion of the scalar part, now we can go through the tensor part. In terms of the number  of e-folds  the  tensor power spectrum, by means of Eqs.(\ref{Lag}) and (\ref{PTphi}),  is given by
\begin{equation}
\label{Pt}
{{\cal P}_t}=\frac{{2V_0{\chi ^2}{{3{M_p}^{-4}{\pi^{-2}}}}}}{\Bigg[{{\Big{(}\frac{N}{\Lambda } + {{({{({{(\frac{{{2^{ - \alpha }}{3^{ - 1 + \alpha }}{M^{4( - 1 + \alpha )}}V_0^{1 - \alpha }{{({M_p}s)}^{2\alpha }}}}{\alpha })}^{\frac{1}{{1 - 2\alpha }}}}\chi )}^{\frac{{1 - 2\alpha }}{{s( - 1 + \alpha ) + 2\alpha }}}})}^{s + \frac{{2 - f}}{{ - 2 + f + 2\alpha }}}}\Big{)}}^{\frac{1}{{s + \frac{{2 - f}}{{ - 2 + f + 2\alpha }}}}}}\Bigg]^{-s}}.
\end{equation}
The tensor spectral index is defined as
\begin{equation}
\label{ntdef}
{n_t} \equiv \frac{{d\ln
{{\cal P}_t}}}{{d\ln k}}.
\end{equation}
Now by using Eqs.(\ref{exithorizon}),  (\ref{Pt}), and the above equation one can obtain
\begin{eqnarray}
\label{ntn}
{n_t} &=& \frac{{-s( - 2 + f + 2\alpha )}}{{((2 - f)(1 - s) + 2\alpha s)\Lambda }}\cr
&\times&\frac{1}{{\frac{{N}}{\Lambda } + {{\Big{(}{{({{(\frac{{{2^{ - \alpha }}{3^{ - 1 + \alpha }}{M^{4( - 1 + \alpha )}}V_0^{1 - \alpha }{{({M_p}s)}^{2\alpha }}}}{\alpha })}^{\frac{1}{{1 - 2\alpha }}}}\chi )}^{\frac{{1 - 2\alpha }}{{s( - 1 + \alpha ) + 2\alpha }}}}\Big{)}}^{s + \frac{{2 - f}}{{ - 2 + f + 2\alpha }}}}}}.
\end{eqnarray}
To measure the amplitude of the primordial fluctuations  we need to calculate the tensor-to-scalar ratio which is defined as
\begin{equation}
\label{rdef}
r \equiv \frac{{{{\cal P}_t}}}{{{{\cal P}_s}}},
\end{equation}
where by using  Eqs. (\ref{Psk}), (\ref{Pt})  and (\ref{rdef}),  it can be expressed by
\begin{eqnarray}
\label{rn}
r& =& {\Bigg[{\Big[\frac{N}{\Lambda } + {\Big{(}{({(\frac{{{2^{ - \alpha }}{3^{ - 1 + \alpha }}{M^{4( - 1 + \alpha )}}V_0^{1 - \alpha }{{({M_p}s)}^{2\alpha }}}}{\alpha })^{\frac{1}{{1 - 2\alpha }}}}\chi )^{\frac{{1 - 2\alpha }}{{s( - 1 + \alpha ) + 2\alpha }}}}\Big{)}^{s + \frac{{2 - f}}{{ - 2 + f + 2\alpha }}}}\Big]^{\frac{1}{{s + \frac{{2 - f}}{{ - 2 + f + 2\alpha }}}}}}\Bigg]^{s + \frac{{(4 - 6f)\alpha }}{{ - 2 + f + 2\alpha }}}}\cr
&\times&\frac{{48{c_{s}}{V_0}{{({6^\alpha }M_{_p}^{^{8 - 14\alpha }}V_0^{^{ - 2 + 3\alpha }}{{(s)}^{ - 2\alpha }}\alpha {\mu ^{4 - 4\alpha }})}^{\frac{1}{{1 - 2\alpha }}}}}}{{M_p^4\chi }}.
\end{eqnarray}
The consistency relation between  observable $r$ and $n_t$ in the non-canonical inflation is as follows
\begin{equation}
\label{rnt3}
r \approx -8 c_s n_t,
\end{equation}
 which has an extra $c_s$ coefficient comparing to the canonical case, i.e. ($r=-8n_t$) \cite{Unn12,Unn13}.
Replacing  Eq.(\ref{ntn}) into Eq(\ref{rnt3}) gives
\begin{eqnarray}
\label{rnt2}
r &\approx & \frac{{8 c_s s( - 2 + f + 2\alpha )}}{{((2 - f)(1 - s) + 2\alpha s)\Lambda }}\cr
&\times&\frac{1}{{\frac{{N}}{\Lambda } + {{\Big{(}{{({{(\frac{{{2^{ - \alpha }}{3^{ - 1 + \alpha }}{M^{4( - 1 + \alpha )}}V_0^{1 - \alpha }{{({M_p}s)}^{2\alpha }}}}{\alpha })}^{\frac{1}{{1 - 2\alpha }}}}\chi )}^{\frac{{1 - 2\alpha }}{{s( - 1 + \alpha ) + 2\alpha }}}}\Big{)}}^{s + \frac{{2 - f}}{{ - 2 + f + 2\alpha }}}}}}.
\end{eqnarray}
Subsequently we are going to check the accuracy and consistency of our theoretical results. To do so we have to make a comparison with observation.   One of the best criterions for our aim could be considered is the data risen by Planck $2013$ and $2015$ \cite{Planck2015}. It is distinct that one of the most important results of Planck data is the  $r-n_s$ diagram and the validity of different models relies on their compatibility to this observation.
Therefore,  by virtue of
Eqs.(\ref{nsk}) and (\ref{rnt2}) we excited to depict the $r-n_s$ diagram for our scenario. This diagram is shown in figure \ref{fignsr}. Besides, the marginalized likelihoods based on Confidence Levels (CLs)  68\% and 95\% are allowed by
Planck  $2015$ ,  TT, TE, EE+lowP data \cite{Planck2015} and we illustrated them in the figure \ref{fignsr}. Predictions of our model are specified by solid black line for the values of $\alpha=3$, $\kappa=3.02\times 10^{-12}$,  $f=10^{-4}$ and $\lambda=3.5$. From figure \ref{fignsr} it could be visualized that our results  can be considered in acceptable ranges compared to the observations. Then, it could be concluded this scenario is able to  be regarded as a valid case for explaining the inflationary scenario.
 It is clear that
the grey, red and blue  CLs are correspond to
Planck $2013$, Planck $2015$ TT+lowP and Planck $2015$ TT, TE, EE+lowP data
respectively \cite{Planck2015}.
Also in Table (\ref{FT}) we want to study the behaviour of the  parameter $\lambda$, i.e. the effect of low anisotropy; one can obviously observe that the  prediction of the model for
perturbation parameters are relied upon the specific values of free
parameters $f,~\alpha, ~\kappa $ but  different values for  parameter $\lambda$. Given the Table (\ref{FT}), it is clear that  the $r-n_s$ diagram   for $\alpha=3$ and $\lambda=3.5$ is in more consistency with the Planck  $2015$  TT, TE, EE+lowP results \cite{Planck2015}. Meanwhile, at $\lambda=1$ and $\alpha=3$ the results lead to the non-canonical but  isotropic Universe, that the $\mathcal{P}_s$has a little bit deviation of its observed value by Planck $\mathcal{P}^{\ast}_s=2.17\times 10^{-9}$

\begin{figure}[ht]
\centering
\includegraphics[scale=.50]{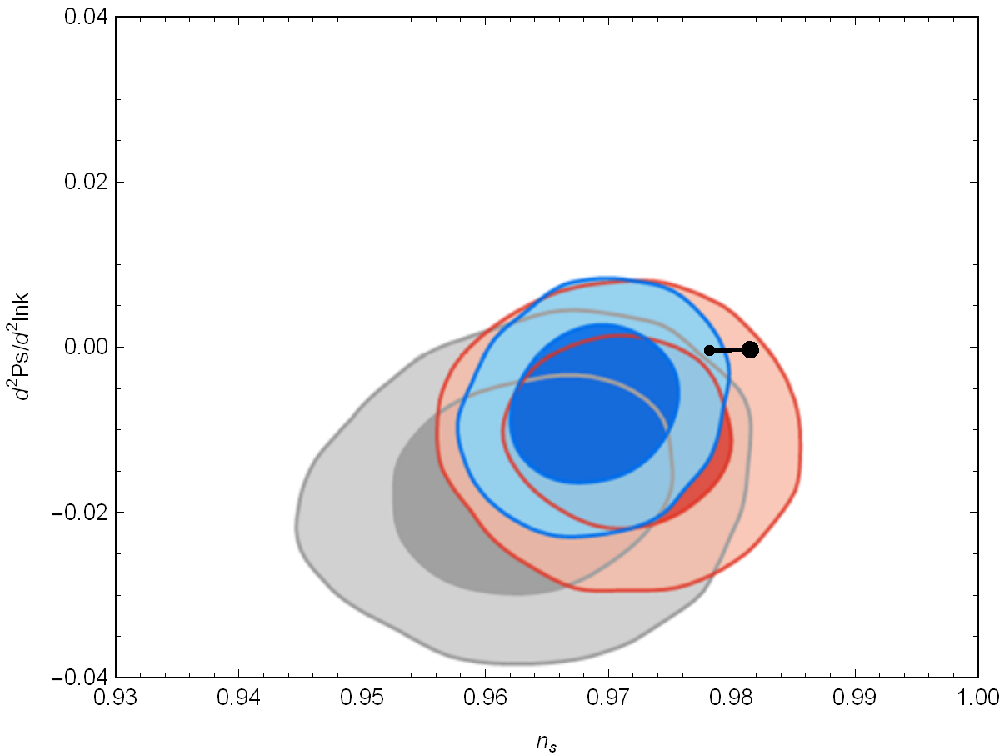}
\caption{{\it{
 The $d{n_s}/dN - {n_s}$ diagram show  Prediction of
the non-canonical intermediate inflationary model in anisotropy background for the specified values of  $\alpha=3$, $\kappa=3.02\times 10^{-12}$,  $f=10^{-4}$ ,$M_p=10^{18}$, $M=10^{12}$ and $\lambda=3.5$,
in comparison with the observational results of Planck 2015. Where the likelihood  of  Planck  $2013$  (grey contours), Planck  $2015$  TT+lowP(red contours), Planck  $2015$  TT,TE,EE+lowP(blue contours) and   the thick black line
indicate the predictions of our model in which small and large dots are the amount of $n_s$ at the  e-folding value  $N=55,~N=65$.}}}
\label{fignsdns1}
\end{figure}

\begin{table}[h]
  \centering
  {\footnotesize
  \begin{tabular}{p{1.2cm}p{1.5cm}p{0.8cm}p{0.8cm}p{1.2cm}p{2.5cm}p{2.5cm}}
    \hline
              $ \alpha$ & $\kappa*10^{-12}$ & $\ \ f$ & $\lambda$ &  $\ \ n_s$   &\ \ $ r$  & $\qquad \mathcal{P}_s$    \\[0.1mm]
         $3$   & $3.02$ & $10^{-4}$ &  $3.5$ &  $0.978$   & $0.078$ & $2.17\times 10^{-9}$   \\[2mm]
         $3$   & $3.02$ & $10^{-4}$ &  $1$ &  $0.978$   & $0.078$ & $6.199\times 10^{-10}$   \\[2mm]
          $1$  & $3.02$ &  $10^{-4}$ &  $3.5$   & $0.964$  & $0.167$ & $2.7\times 10^{590950}$    \\[0.1mm]
         $2$  & $3.02$ &  $10^{-4}$ &  $2.5$  & $0.976$ & $0.290$ & $7293.78$   \\[0.1mm]
         $1$  & $3.02$ &  $10^{-4}$ &  $1$  & $0.964$ & $0.130$ & $9.07\times 10^{587695}$   \\[0.1mm]
  \end{tabular}
  }
  \caption{\footnotesize The  prediction of the model for the perturbation parameters $n_s$, $r$ and $\mathcal{P}_s$ are prepared for  different values of the free parameters $\lambda$ and $ \alpha$ besides the specified values of other free parameters. Also we used $M_p=10^{18}$, $M=10^{12}$ and $N=55$.  This analyze shows the best behaviour for the first row of the table. On the second row we can see the behaviour of non-canonical model in isotropic background in which there is some deviations almost around one order for $\mathcal{P}_s$. We also examine the canonical case with anisotropic condition on third row and the result was very far from the observed results, especially the value for $\mathcal{P}_s$. The amounts for free parameters on the fourth row are supplied for more clarity in comparison. And finally we consider the canonical case in an isotropic background and the results again were not according to excepted results originated from observations}\label{FT}
\end{table}
Now we can turn into the running spectral index,  $\alpha_s=d{n_s}/dN -{n_s}$, behaviour  in comparison to the observational results originated of Planck data. So at first we regard   $\alpha=3$, $\kappa=3.02\times 10^{-12}$,  $f=10^{-4}$ ,$M_p=10^{18}$, $M=10^{12}$, $\mu=10^{-6}$, and $\lambda=3.5$. Then, by using  Eqs. (\ref{nsk}) and
(\ref{dnsk}) we will plot $d{n_s}/dN$ versus $n_s$.  The plot \ref{fignsdns1} shows the prediction of the model could lie insides the joint 68\% CLs region of Planck  $2015$  TT, TE, EE+lowP data, and satisfies the agreement with observations  \cite{Planck2015}.

\newpage

\section{Conclusions}\label{seccon}
A well-known class of scale factors namely the intermediate ones for a non-canonical Lagrangian in the anisotropic background has been investigated. The main motivation of doing such investigation goes back to cope with drawbacks of canonical and isotropic version of inflationary scenarios. Despite some complications in formulas and calculations, have been raised because of extension in the model, fortunately and without any hand made conditions it has bee shown that the obtained potential automatically takes a steep form, i.e. $V = V_0 \: \phi^{s}$ with $s<0$.
This class of potentials, as it has been shown, can be considered as a suitable candidate to run the inflation in an acceptable process, compared to observational constraints. To examine our proposal we have been followed the slow-roll method and all necessary parameters have been estimated based on a powerful criterion such as Planck $2015$. Amongst those aforementioned observables we have been focused on the amplitudes of scalar and tensor perturbations, their ratio, scalar and tensor spectral indices and their running as well. So by combining resulted potential and slow-rolling approach we have been tried to examine the accuracy of our estimations and also the claims about the succusses of Non-canonical anisotropic model. It has been clear one of the most important results of Planck data is the $r-n_s$ diagram, and  the validity of theoretical models rely upon their acceptable compatibility with this criterion. Whereas we have been obtained all the necessary instruments to examine the validity of our results  we could plot  them based on the original figures originated from the Planck collaboration papers, e.g. \cite{Planck2015}. Therefore based on our investigations for  $r-n_s$ analyzes at first we have been supplied a diagram in Fig.\ref{fignsr} and it has been observed that non-canonical anisotropic inflation with an intermediate scalar field could be considered as a suitable candidate to drive inflation. And then, consequently  in Table. \ref{FT} different asymptotical behaviour based on definitions for the Lagrangian and also BI metric and the comparability with $n_s$ and $\mathcal{P}_s$  appeared in Planck have been evaluated. The best free parameters have been obtained as $\alpha=3$,  $\lambda=3.5$ and  $f=10^{-4}$ in which we have been used some specific values for other parameters like the $\kappa=3.02\times 10^{-12}$ ,$M_p=10^{18}$, and $M=10^{12}$. To visualize the aforementioned asymptotical behaviour at first we have been considered  $\lambda=1$ to go back to isotropic background. From this point of view it has been concluded that even for the well accepted non-canonical Lagrangian the results in the isotropic universe have some deviations compared to data. Even more the situation could be absolutely teerible for canonical Lagrangian even in anisotropic background. Besides, in Fig.\ref{fignsdns1}, the predictions for the running spectral index have been appeared also in acceptable ranges comparing with observational data, that has been relied  insides the joint 68\% CLs region of Planck 2015 TT, TE, EE+lowP data \cite{Planck2015}.

\section{Acknowledgement}

The authors thank  the anonymous referee for his/her useful comments and suggestions have resulted in an improved version of their manuscript. HS would like to appreciate IPM, and specially H. Firouzjahi, for their hospitality and constructive discussions during his visit of there. Also he is grateful ICTP, during Summer School 2018, to give him constructive ideas about inflation and primordial fluctuations. He appreciates G. Ellis, A. Weltamn and UCT to arrange his short visit of there and good discussions about primordial universe. He is also grateful his wife E. Avirdi for her valuable notes and being patience during our stay in South Africa.


\begin{thebibliography}{99}

% Inflationary models
\bibitem{Sta80}    A.~A.~Starobinsky,
  ``A New Type of Isotropic Cosmological Models Without Singularity,''
  Phys.\ Lett.\ B {\bf 91}, 99 (1980)
  [Phys.\ Lett.\  {\bf 91B}, 99 (1980)]
  [Adv.\ Ser.\ Astrophys.\ Cosmol.\  {\bf 3}, 130 (1987)].
  doi:10.1016/0370-2693(80)90670-X
  %%CITATION = doi:10.1016/0370-2693(80)90670-X;%%
  %3870 citations counted in INSPIRE as of 06 Dec 2018
\bibitem{Gut81}    A.~H.~Guth,
  ``The Inflationary Universe: A Possible Solution to the Horizon and Flatness Problems,''
  Phys.\ Rev.\ D {\bf 23}, 347 (1981)
  [Adv.\ Ser.\ Astrophys.\ Cosmol.\  {\bf 3}, 139 (1987)].
  doi:10.1103/PhysRevD.23.347
  %%CITATION = doi:10.1103/PhysRevD.23.347;%%
  %7086 citations counted in INSPIRE as of 06 Dec 2018


\bibitem{Lin82} %\cite{Linde:1981mu}
%\bibitem{Linde:1981mu}
  A.~D.~Linde,
  ``A New Inflationary Universe Scenario: A Possible Solution of the Horizon, Flatness, Homogeneity, Isotropy and Primordial Monopole Problems,''
  Phys.\ Lett.\  {\bf 108B}, 389 (1982)
  [Adv.\ Ser.\ Astrophys.\ Cosmol.\  {\bf 3}, 149 (1987)].
  doi:10.1016/0370-2693(82)91219-9
  %%CITATION = doi:10.1016/0370-2693(82)91219-9;%%
  %4394 citations counted in INSPIRE as of 06 Dec 2018


\bibitem{Alb} %\cite{Albrecht:1982wi}
%\bibitem{Albrecht:1982wi}
  A.~Albrecht and P.~J.~Steinhardt,
 ``Cosmology for Grand Unified Theories with Radiatively Induced Symmetry Breaking,''
  Phys.\ Rev.\ Lett.\  {\bf 48}, 1220 (1982)
  [Adv.\ Ser.\ Astrophys.\ Cosmol.\  {\bf 3}, 158 (1987)].
  doi:10.1103/PhysRevLett.48.1220
  %%CITATION = doi:10.1103/PhysRevLett.48.1220;%%
  %3856 citations counted in INSPIRE as of 06 Dec 2018


\bibitem{Lin83} %\cite{Linde:1983gd}
%\bibitem{Linde:1983gd}
  A.~D.~Linde,
``Chaotic Inflation,''
  Phys.\ Lett.\  {\bf 129B}, 177 (1983).
  doi:10.1016/0370-2693(83)90837-7
  %%CITATION = doi:10.1016/0370-2693(83)90837-7;%%
  %2676 citations counted in INSPIRE as of 06 Dec 2018


\bibitem{Lin86a} %\cite{Linde:1986fd}
%\bibitem{Linde:1986fd}
  A.~D.~Linde,
``Eternally Existing Selfreproducing Chaotic Inflationary Universe,''
  Phys.\ Lett.\ B {\bf 175}, 395 (1986).
  doi:10.1016/0370-2693(86)90611-8
  %%CITATION = doi:10.1016/0370-2693(86)90611-8;%%
  %680 citations counted in INSPIRE as of 06 Dec 2018


\bibitem{Lin86b} %\cite{Linde:1986fc}
%\bibitem{Linde:1986fc}
  A.~D.~Linde,
 ``Eternal Chaotic Inflation,''
  Mod.\ Phys.\ Lett.\ A {\bf 1}, 81 (1986).
  doi:10.1142/S0217732386000129
  %%CITATION = doi:10.1142/S0217732386000129;%%
  %265 citations counted in INSPIRE as of 06 Dec 2018
% Reviews on inflation


\bibitem{Lid00} A. R. Liddle, D. H. Lyth, \emph{Cosmological Inflation and Large-scale Structure}, Cambridge University Press (2000).


\bibitem{Bas} %\cite{Bassett:2005xm}
%\bibitem{Bassett:2005xm}
  B.~A.~Bassett, S.~Tsujikawa and D.~Wands,
  ``Inflation dynamics and reheating,''
  Rev.\ Mod.\ Phys.\  {\bf 78}, 537 (2006)
  doi:10.1103/RevModPhys.78.537
  [astro-ph/0507632].
  %%CITATION = doi:10.1103/RevModPhys.78.537;%%
  %634 citations counted in INSPIRE as of 06 Dec 2018


\bibitem{Lem} M. Lemoine, J. Martin, P. Peter, \emph{Inflationary Cosmology}, Lect. Notes Phys. 738 (Springer, Berlin Heidelberg 2008).



\bibitem{Kin} %\cite{Kinney:2009vz}
%\bibitem{Kinney:2009vz}
  W.~H.~Kinney,
  ``TASI Lectures on Inflation,''
  arXiv:0902.1529 [astro-ph.CO].
  %%CITATION = ARXIV:0902.1529;%%
  %72 citations counted in INSPIRE as of 06 Dec 2018


\bibitem{Bau09} %\cite{Baumann:2009ds}
%\bibitem{Baumann:2009ds}
  D.~Baumann,
  ``Inflation,''
  arXiv:0907.5424 [hep-th].
  %%CITATION = doi:10.1142/9789814327183_0010;%%
  %555 citations counted in INSPIRE as of 06 Dec 2018


\bibitem{Bau14} %\cite{Baumann:2014nda}
%\bibitem{Baumann:2014nda}
  D.~Baumann and L.~McAllister,
  ``Inflation and String Theory,''
  doi:10.1017/CBO9781316105733
  arXiv:1404.2601 [hep-th].
  %%CITATION = doi:10.1017/CBO9781316105733;%%
  %304 citations counted in INSPIRE as of 06 Dec 2018


\bibitem{Liddle0} %\cite{Liddle:1999mq}
%\bibitem{Liddle:1999mq}
  A.~R.~Liddle,
  ``An Introduction to cosmological inflation,''
  astro-ph/9901124.
  %%CITATION = ASTRO-PH/9901124;%%
  %77 citations counted in INSPIRE as of 06 Dec 2018




\bibitem{Langl}%\cite{Langlois:2004de}
%\bibitem{Langlois:2004de}
  D.~Langlois,
  ``Inflation, quantum fluctuations and cosmological perturbations,''
  hep-th/0405053.
  %%CITATION = HEP-TH/0405053;%%
  %60 citations counted in INSPIRE as of 06 Dec 2018


\bibitem{Lyth} %\cite{Lyth:1998xn}
%\bibitem{Lyth:1998xn}
  D.~H.~Lyth and A.~Riotto,
  ``Particle physics models of inflation and the cosmological density perturbation,''
  Phys.\ Rept.\  {\bf 314}, 1 (1999)
  doi:10.1016/S0370-1573(98)00128-8
  [hep-ph/9807278].
  %%CITATION = doi:10.1016/S0370-1573(98)00128-8;%%
  %1675 citations counted in INSPIRE as of 06 Dec 2018



\bibitem{Guth00} %\cite{Guth:2000ka}
%\bibitem{Guth:2000ka}
  A.~H.~Guth,
  ``Inflation and eternal inflation,''
  Phys.\ Rept.\  {\bf 333}, 555 (2000)
  doi:10.1016/S0370-1573(00)00037-5
  [astro-ph/0002156].
  %%CITATION = doi:10.1016/S0370-1573(00)00037-5;%%
  %229 citations counted in INSPIRE as of 06 Dec 2018


\bibitem{Lidsey97}%\cite{Lidsey:1995np}
%\bibitem{Lidsey:1995np}
  J.~E.~Lidsey, A.~R.~Liddle, E.~W.~Kolb, E.~J.~Copeland, T.~Barreiro and M.~Abney,
  ``Reconstructing the inflation potential : An overview,''
  Rev.\ Mod.\ Phys.\  {\bf 69}, 373 (1997)
  doi:10.1103/RevModPhys.69.373
  [astro-ph/9508078].
  %%CITATION = doi:10.1103/RevModPhys.69.373;%%


\bibitem{Mukhanov-etal} %\cite{Mukhanov:1990me}
%\bibitem{Mukhanov:1990me}
  V.~F.~Mukhanov, H.~A.~Feldman and R.~H.~Brandenberger,
  ``Theory of cosmological perturbations. Part 1. Classical perturbations. Part 2. Quantum theory of perturbations. Part 3. Extensions,''
  Phys.\ Rept.\  {\bf 215}, 203 (1992).
  doi:10.1016/0370-1573(92)90044-Z
  %%CITATION = doi:10.1016/0370-1573(92)90044-Z;%%
  %2693 citations counted in INSPIRE as of 06 Dec 2018


\bibitem{Haidar} Haidar Sheikhahmadi,   ``Schwinger-Keldysh mechanism in extended quasi single field inflation,'' To be appeared in EPJC,
[arXiv:1901.01905 [gr-qc]].

\bibitem{Haidar2} %\cite{Sheikhahmadi:2016wyz}
%\bibitem{Sheikhahmadi:2016wyz}
  H.~Sheikhahmadi, E.~N.~Saridakis, A.~Aghamohammadi and K.~Saaidi,
  ``Hamilton-Jacobi formalism for inflation with non-minimal derivative coupling,''
  JCAP {\bf 1610}, no. 10, 021 (2016)
  doi:10.1088/1475-7516/2016/10/021
  [arXiv:1603.03883 [gr-qc]].
  %%CITATION = doi:10.1088/1475-7516/2016/10/021;%%
  %15 citations counted in INSPIRE as of 06 Dec 2018





\bibitem{Arm} %\cite{ArmendarizPicon:1999rj}
%\bibitem{ArmendarizPicon:1999rj}
  C.~Armendariz-Picon, T.~Damour and V.~F.~Mukhanov,
  ``k - inflation,''
  Phys.\ Lett.\ B {\bf 458}, 209 (1999)
  doi:10.1016/S0370-2693(99)00603-6
  [hep-th/9904075].
  %%CITATION = doi:10.1016/S0370-2693(99)00603-6;%%
  %1358 citations counted in INSPIRE as of 06 Dec 2018


\bibitem{Gar} %\cite{Garriga:1999vw}
%\bibitem{Garriga:1999vw}
  J.~Garriga and V.~F.~Mukhanov,
 ``Perturbations in k-inflation,''
  Phys.\ Lett.\ B {\bf 458}, 219 (1999)
  doi:10.1016/S0370-2693(99)00602-4
  [hep-th/9904176].
  %%CITATION = doi:10.1016/S0370-2693(99)00602-4;%%
  %874 citations counted in INSPIRE as of 06 Dec 2018


\bibitem{Li}%\cite{Li:2012vta}
%\bibitem{Li:2012vta}
  S.~Li and A.~R.~Liddle,
 ``Observational constraints on K-inflation models,''
  JCAP {\bf 1210}, 011 (2012)
  doi:10.1088/1475-7516/2012/10/011
  [arXiv:1204.6214 [astro-ph.CO]].
  %%CITATION = doi:10.1088/1475-7516/2012/10/011;%%
  %12 citations counted in INSPIRE as of 06 Dec 2018


\bibitem{Hwa} %\cite{Hwang:2002fp}
%\bibitem{Hwang:2002fp}
  J.~c.~Hwang and H.~Noh,
  ``Cosmological perturbations in a generalized gravity including tachyonic condensation,''
  Phys.\ Rev.\ D {\bf 66}, 084009 (2002)
  doi:10.1103/PhysRevD.66.084009
  [hep-th/0206100].
  %%CITATION = doi:10.1103/PhysRevD.66.084009;%%
  %111 citations counted in INSPIRE as of 06 Dec 2018


\bibitem{Fra10a} %\cite{Franche:2009gk}
%\bibitem{Franche:2009gk}
  P.~Franche, R.~Gwyn, B.~Underwood and A.~Wissanji,
``Attractive Lagrangians for Non-Canonical Inflation,''
  Phys.\ Rev.\ D {\bf 81}, 123526 (2010)
  doi:10.1103/PhysRevD.81.123526
  [arXiv:0912.1857 [hep-th]].
  %%CITATION = doi:10.1103/PhysRevD.81.123526;%%
  %34 citations counted in INSPIRE as of 06 Dec 2018



\bibitem{Fra10b} %\cite{Franche:2010yj}
%\bibitem{Franche:2010yj}
  P.~Franche, R.~Gwyn, B.~Underwood and A.~Wissanji,
``Initial Conditions for Non-Canonical Inflation,''
  Phys.\ Rev.\ D {\bf 82}, 063528 (2010)
  doi:10.1103/PhysRevD.82.063528
  [arXiv:1002.2639 [hep-th]].
  %%CITATION = doi:10.1103/PhysRevD.82.063528;%%
  %12 citations counted in INSPIRE as of 06 Dec 2018

\bibitem{Unn12} %\cite{Unnikrishnan:2012zu}
%\bibitem{Unnikrishnan:2012zu}
  S.~Unnikrishnan, V.~Sahni and A.~Toporensky,
  ``Refining inflation using non-canonical scalars,''
  JCAP {\bf 1208}, 018 (2012)
  doi:10.1088/1475-7516/2012/08/018
  [arXiv:1205.0786 [astro-ph.CO]].
  %%CITATION = doi:10.1088/1475-7516/2012/08/018;%%
  %38 citations counted in INSPIRE as of 06 Dec 2018




\bibitem{Unn13} %\cite{Unnikrishnan:2013vga}
%\bibitem{Unnikrishnan:2013vga}
  S.~Unnikrishnan and V.~Sahni,
 ``Resurrecting power law inflation in the light of Planck results,''
  JCAP {\bf 1310}, 063 (2013)
  doi:10.1088/1475-7516/2013/10/063
  [arXiv:1305.5260 [astro-ph.CO]].
  %%CITATION = doi:10.1088/1475-7516/2013/10/063;%%
  %28 citations counted in INSPIRE as of 06 Dec 2018


\bibitem{Zha14a} %\cite{Zhang:2014dja}
%\bibitem{Zhang:2014dja}
  X.~M.~Zhang and j.~Y.~Zhu,
  ``Extension of warm inflation to noncanonical scalar fields,''
  Phys.\ Rev.\ D {\bf 90}, no. 12, 123519 (2014)
  doi:10.1103/PhysRevD.90.123519
  [arXiv:1402.0205 [gr-qc]].
  %%CITATION = doi:10.1103/PhysRevD.90.123519;%%
  %16 citations counted in INSPIRE as of 06 Dec 2018


\bibitem{Gol} %\cite{Golanbari:2014lca}
%\bibitem{Golanbari:2014lca}
  T.~Golanbari, A.~Mohammadi and K.~Saaidi,
  ``Brane inflation driven by noncanonical scalar field,''
  Phys.\ Rev.\ D {\bf 89}, no. 10, 103529 (2014)
  doi:10.1103/PhysRevD.89.103529
  [arXiv:1405.6359 [astro-ph.CO]].
  %%CITATION = doi:10.1103/PhysRevD.89.103529;%%
  %12 citations counted in INSPIRE as of 06 Dec 2018


\bibitem{Nazavari} %\cite{Nazavari:2016yaa}
%\bibitem{Nazavari:2016yaa}
  N.~Nazavari, A.~Mohammadi, Z.~Ossoulian and K.~Saaidi,
``Intermediate inflation driven by DBI scalar field,''
  Phys.\ Rev.\ D {\bf 93}, no. 12, 123504 (2016)
  doi:10.1103/PhysRevD.93.123504
  [arXiv:1708.03676 [gr-qc]].
  %%CITATION = doi:10.1103/PhysRevD.93.123504;%%
  %7 citations counted in INSPIRE as of 06 Dec 2018


\bibitem{Mar13}%\cite{Martin:2013tda}
%\bibitem{Martin:2013tda}
  J.~Martin, C.~Ringeval and V.~Vennin,
  ``Encyclop{\ae}dia Inflationaris,''
  Phys.\ Dark Univ.\  {\bf 5-6}, 75 (2014)
  doi:10.1016/j.dark.2014.01.003
  [arXiv:1303.3787 [astro-ph.CO]].
  %%CITATION = doi:10.1016/j.dark.2014.01.003;%%
  %476 citations counted in INSPIRE as of 06 Dec 2018.



\bibitem{Mar14} %\cite{Martin:2013nzq}
%\bibitem{Martin:2013nzq}
  J.~Martin, C.~Ringeval, R.~Trotta and V.~Vennin,
  ``The Best Inflationary Models After Planck,''
  JCAP {\bf 1403}, 039 (2014)
  doi:10.1088/1475-7516/2014/03/039
  [arXiv:1312.3529 [astro-ph.CO]].
  %%CITATION = doi:10.1088/1475-7516/2014/03/039;%%
  %184 citations counted in INSPIRE as of 06 Dec 2018


\bibitem{Hos14a} %\cite{Hossain:2014coa}
%\bibitem{Hossain:2014coa}
  M.~W.~Hossain, R.~Myrzakulov, M.~Sami and E.~N.~Saridakis,
  ``Class of quintessential inflation models with parameter space consistent with BICEP2,''
  Phys.\ Rev.\ D {\bf 89}, no. 12, 123513 (2014)
  doi:10.1103/PhysRevD.89.123513
  [arXiv:1404.1445 [gr-qc]].
  %%CITATION = doi:10.1103/PhysRevD.89.123513;%%
  %35 citations counted in INSPIRE as of 06 Dec 2018


\bibitem{Barrow11} %\cite{Barrow:1990vx}
%\bibitem{Barrow:1990vx}
  J.~D.~Barrow,
  ``Graduated Inflationary Universes,''
  Phys.\ Lett.\ B {\bf 235}, 40 (1990).
  doi:10.1016/0370-2693(90)90093-L
  %%CITATION = doi:10.1016/0370-2693(90)90093-L;%%
  %281 citations counted in INSPIRE as of 06 Dec 2018


\bibitem{Vallinotto} %\cite{Vallinotto:2003vf}
%\bibitem{Vallinotto:2003vf}
  A.~Vallinotto, E.~J.~Copeland, E.~W.~Kolb, A.~R.~Liddle and D.~A.~Steer,
  ``Inflationary potentials yielding constant scalar perturbation spectral indices,''
  Phys.\ Rev.\ D {\bf 69}, 103519 (2004)
  doi:10.1103/PhysRevD.69.103519
  [astro-ph/0311005].
  %%CITATION = doi:10.1103/PhysRevD.69.103519;%%
  %40 citations counted in INSPIRE as of 06 Dec 2018




\bibitem{Starobinsky} %\cite{Starobinsky:2005ab}
%\bibitem{Starobinsky:2005ab}
  A.~A.~Starobinsky,
  ``Inflaton field potential producing the exactly flat spectrum of adiabatic perturbations,''
  JETP Lett.\  {\bf 82}, 169 (2005)
  [Pisma Zh.\ Eksp.\ Teor.\ Fiz.\  {\bf 82}, 187 (2005)]
  doi:10.1134/1.2121807
  [astro-ph/0507193].
  %%CITATION = doi:10.1134/1.2121807;%%
  %56 citations counted in INSPIRE as of 06 Dec 2018



\bibitem{Rendall} %\cite{Rendall:2005if}
%\bibitem{Rendall:2005if}
  A.~D.~Rendall,
 ``Intermediate inflation and the slow-roll approximation,''
  Class.\ Quant.\ Grav.\  {\bf 22}, 1655 (2005)
  doi:10.1088/0264-9381/22/9/013
  [gr-qc/0501072].
  %%CITATION = doi:10.1088/0264-9381/22/9/013;%%
  %79 citations counted in INSPIRE as of 06 Dec 2018



\bibitem{Barrow-etal} %\cite{Barrow:1990td}
%\bibitem{Barrow:1990td}
  J.~D.~Barrow and P.~Saich,
``The Behavior of intermediate inflationary universes,''
  Phys.\ Lett.\ B {\bf 249}, 406 (1990).
  doi:10.1016/0370-2693(90)91007-X
  %%CITATION = doi:10.1016/0370-2693(90)91007-X;%%
  %114 citations counted in INSPIRE as of 06 Dec 2018

\bibitem{BarrowNunes}%\cite{Barrow:2007zr}
%\bibitem{Barrow:2007zr}
  J.~D.~Barrow and N.~J.~Nunes,
  ``Dynamics of Logamediate Inflation,''
  Phys.\ Rev.\ D {\bf 76}, 043501 (2007)
  doi:10.1103/PhysRevD.76.043501
  [arXiv:0705.4426 [astro-ph]].
  %%CITATION = doi:10.1103/PhysRevD.76.043501;%%
  %65 citations counted in INSPIRE as of 06 Dec 2018

\bibitem{Muslimov} %\cite{Muslimov:1990be}
%\bibitem{Muslimov:1990be}
  A.~G.~Muslimov,
``On the Scalar Field Dynamics in a Spatially Flat Friedman Universe,''
  Class.\ Quant.\ Grav.\  {\bf 7}, 231 (1990).
  doi:10.1088/0264-9381/7/2/015
  %%CITATION = doi:10.1088/0264-9381/7/2/015;%%
  %175 citations counted in INSPIRE as of 06 Dec 2018


\bibitem{BarowLiddle02}%\cite{Barrow:2006dh}
%\bibitem{Barrow:2006dh}
  J.~D.~Barrow, A.~R.~Liddle and C.~Pahud,
 ``Intermediate inflation in light of the three-year WMAP observations,''
  Phys.\ Rev.\ D {\bf 74}, 127305 (2006)
  doi:10.1103/PhysRevD.74.127305
  [astro-ph/0610807].
  %%CITATION = doi:10.1103/PhysRevD.74.127305;%%
  %77 citations counted in INSPIRE as of 06 Dec 2018

\bibitem{BarrowLiddle}%\cite{Barrow:1993zq}
%\bibitem{Barrow:1993zq}
  J.~D.~Barrow and A.~R.~Liddle,
  ``Perturbation spectra from intermediate inflation,''
  Phys.\ Rev.\ D {\bf 47}, no. 12, R5219 (1993)
  doi:10.1103/PhysRevD.47.R5219
  [astro-ph/9303011].
  %%CITATION = doi:10.1103/PhysRevD.47.R5219;%%
  %144 citations counted in INSPIRE as of 06 Dec 2018
\
\bibitem{kk} %\cite{Rezazadeh:2014fwa}
%\bibitem{Rezazadeh:2014fwa}
  K.~Rezazadeh, K.~Karami and P.~Karimi,
  ``Intermediate inflation from a non-canonical scalar field,''
  JCAP {\bf 1509}, no. 09, 053 (2015)
  doi:10.1088/1475-7516/2015/09/053
  [arXiv:1411.7302 [gr-qc]].
  %%CITATION = doi:10.1088/1475-7516/2015/09/053;%%
  %17 citations counted in INSPIRE as of 06 Dec 2018


\bibitem{mohammadi} %\cite{Mohammadi:2015jka}
%\bibitem{Mohammadi:2015jka}
  A.~Mohammadi, Z.~Ossoulian, T.~Golanbari and K.~Saaidi,
``Intermediate inflation with modified kinetic term,''
  Astrophys.\ Space Sci.\  {\bf 359}, no. 1, 7 (2015).
  doi:10.1007/s10509-015-2458-5
  %%CITATION = doi:10.1007/s10509-015-2458-5;%%
  %6 citations counted in INSPIRE as of 06 Dec 2018


\bibitem{ch1:9} D. N. Spergel, et al., ``{First-Year Wilkinson Microwave Anisotropy Probe (WMAP) Observations: Determination of Cosmological Parameters, Astrophys. J. suppl}'' \textbf{148} 175   (2003).
\bibitem{WM2} G. Hinshaw, et al., ``{Five-Year Wilkinson Microwave Anisotropy Probe (WMAP) Observations: Data Processing, Sky Maps, and Basic Results,  Astrophys. J. Suppl }'' \textbf{180} 225  (2009).


\bibitem{ko} %\cite{Komatsu:2008hk}
%\bibitem{Komatsu:2008hk}
  E.~Komatsu {\it et al.} [WMAP Collaboration],
  ``Five-Year Wilkinson Microwave Anisotropy Probe (WMAP) Observations: Cosmological Interpretation,''
  Astrophys.\ J.\ Suppl.\  {\bf 180}, 330 (2009)
  doi:10.1088/0067-0049/180/2/330
  [arXiv:0803.0547 [astro-ph]].
  %%CITATION = doi:10.1088/0067-0049/180/2/330;%%
  %4718 citations counted in INSPIRE as of 06 Dec 2018


\bibitem{ku} %\cite{Kumar:2010kb}
%\bibitem{Kumar:2010kb}
  S.~Kumar and A.~K.~Yadav,
  ``Some Bianchi Type-V Models of Accelerating Universe with Dark Energy,''
  Mod.\ Phys.\ Lett.\ A {\bf 26}, 647 (2011)
  doi:10.1142/S0217732311035018
  [arXiv:1010.6268 [physics.gen-ph]].
  %%CITATION = doi:10.1142/S0217732311035018;%%
  %56 citations counted in INSPIRE as of 06 Dec 2018


\bibitem{Y1} %\cite{Yadav:2010ah}
%\bibitem{Yadav:2010ah}
  A.~K.~Yadav and L.~Yadav,
  ``Bianchi Type III Anisotropic Dark Energy Model with Constant Deceleration Parameter,''
  Int.\ J.\ Theor.\ Phys.\  {\bf 50}, 218 (2011)
  doi:10.1007/s10773-010-0510-3
  [arXiv:1007.1411 [gr-qc]].
  %%CITATION = doi:10.1007/s10773-010-0510-3;%%
  %74 citations counted in INSPIRE as of 06 Dec 2018




\bibitem{17} %\cite{Saha:2001ig}
%\bibitem{Saha:2001ig}
  B.~Saha,
  ``Spinor field in Bianchi type I universe: Regular solutions,''
  Phys.\ Rev.\ D {\bf 64}, 123501 (2001)
  doi:10.1103/PhysRevD.64.123501
  [gr-qc/0107013].
  %%CITATION = doi:10.1103/PhysRevD.64.123501;%%
  %102 citations counted in INSPIRE as of 06 Dec 2018


\bibitem{18} %\cite{Saha:2003xv}
%\bibitem{Saha:2003xv}
  B.~Saha and T.~Boyadjiev,
  ``Bianchi type I cosmology with scalar and spinor fields,''
  Phys.\ Rev.\ D {\bf 69}, 124010 (2004)
  doi:10.1103/PhysRevD.69.124010
  [gr-qc/0311045].
  %%CITATION = doi:10.1103/PhysRevD.69.124010;%%
  %100 citations counted in INSPIRE as of 06 Dec 2018


\bibitem{aghaohamadi} %\cite{Aghamohammadi:2017emk}
%\bibitem{Aghamohammadi:2017emk}
  A.~Aghamohammadi, H.~Hossienkhani and K.~Saaidi,
  ``Anisotropy effects on baryogenesis in f(R) theories of gravity,''
  Mod.\ Phys.\ Lett.\ A {\bf 33}, no. 13, 1850072 (2018)
  doi:10.1142/S0217732318500724
  [arXiv:1709.06996 [physics.gen-ph]].
  %%CITATION = doi:10.1142/S0217732318500724;%%
  %1 citations counted in INSPIRE as of 06 Dec 2018



\bibitem{r7} E. Kasner,
``An algebraic solution of the Einstein equations,''
Trans. Am. Math. Soc. \textbf{27}, 101  (1925)
doi:10.1090/S0002-9947-1925-1501301-4


\bibitem{r8} %\cite{Barrow:2005dn}
%\bibitem{Barrow:2005dn}
  J.~D.~Barrow and T.~Clifton,
  ``Exact cosmological solutions of scale-invariant gravity theories,''
  Class.\ Quant.\ Grav.\  {\bf 23}, L1 (2006)
  doi:10.1088/0264-9381/23/1/L01
  [gr-qc/0509085].
  %%CITATION = doi:10.1088/0264-9381/23/1/L01;%%
  %74 citations counted in INSPIRE as of 06 Dec 2018



\bibitem{r9}%\cite{Clifton:2006kc}
%\bibitem{Clifton:2006kc}
  T.~Clifton and J.~D.~Barrow,
``Further exact cosmological solutions to higher-order gravity theories,''
  Class.\ Quant.\ Grav.\  {\bf 23}, 2951 (2006)
  doi:10.1088/0264-9381/23/9/011
  [gr-qc/0601118].
  %%CITATION = doi:10.1088/0264-9381/23/9/011;%%
  %75 citations counted in INSPIRE as of 06 Dec 2018


\bibitem{18a} %\cite{Hossienkhani:2017uku}
%\bibitem{Hossienkhani:2017uku}
  H.~Hossienkhani, A.~Aghamohammadi, A.~Jafari, S.~W.~Rabieei and A.~Refaei,
``Effects of low anisotropy on interacting holographic and new agegraphic scalar fields models of dark energy,''
  Phys.\ Dark Univ.\  {\bf 18}, 17 (2017).
  doi:10.1016/j.dark.2017.09.004
  %%CITATION = doi:10.1016/j.dark.2017.09.004;%%
  %3 citations counted in INSPIRE as of 06 Dec 2018




\bibitem{35a} %\cite{Ellis:1998ct}
%\bibitem{Ellis:1998ct}
  G.~F.~R.~Ellis and H.~van Elst,
  ``Cosmological models: Cargese lectures 1998,''
  NATO Sci.\ Ser.\ C {\bf 541}, 1 (1999)
  [gr-qc/9812046].
  %%CITATION = doi:10.1007/978-94-011-4455-1_1;%%
  %341 citations counted in INSPIRE as of 06 Dec 2018



\bibitem{36}  %\cite{Fayaz:2014bja}
%\bibitem{Fayaz:2014bja}
  V.~Fayaz, H.~Hossienkhani, M.~Amirabadi and N.~Azimi,
 ``Anisotropic cosmological models in $\mathcal{f}(R,T)$ gravity according to holographic and new agegraphic dark energy,''
  Astrophys.\ Space Sci.\  {\bf 353}, 301 (2014).
  doi:10.1007/s10509-014-2053-1
  %%CITATION = doi:10.1007/s10509-014-2053-1;%%
  %10 citations counted in INSPIRE as of 06 Dec 2018


\bibitem{37}  %\cite{Hossienkhani:2014cja}
%\bibitem{Hossienkhani:2014cja}
  H.~Hossienkhani, A.~Najafi and N.~Azimi,
 ``Reconstruction of f(R,T) gravity in anisotropic cosmological models of accelerating universe,''
  Astrophys.\ Space Sci.\  {\bf 353}, 311 (2014).
  doi:10.1007/s10509-014-2068-7
  %%CITATION = doi:10.1007/s10509-014-2068-7;%%
  %11 citations counted in INSPIRE as of 06 Dec 2018


\bibitem{38} %\cite{Hossienkhani:2014zoa}
%\bibitem{Hossienkhani:2014zoa}
  H.~Hossienkhani and A.~Pasqua,
 ``Thermal relic abundance and anisotropy due to modified gravity,''
  Astrophys.\ Space Sci.\  {\bf 349}, 39 (2014).
  doi:10.1007/s10509-013-1645-5
  %%CITATION = doi:10.1007/s10509-013-1645-5;%%
  %16 citations counted in INSPIRE as of 06 Dec 2018


\bibitem{Kho2}  %\cite{Khoury:2003aq}
%\bibitem{Khoury:2003aq}
  J.~Khoury and A.~Weltman,
 ``Chameleon fields: Awaiting surprises for tests of gravity in space,''
  Phys.\ Rev.\ Lett.\  {\bf 93}, 171104 (2004)
  doi:10.1103/PhysRevLett.93.171104
  [astro-ph/0309300].
  %%CITATION = doi:10.1103/PhysRevLett.93.171104;%%
  %975 citations counted in INSPIRE as of 06 Dec 2018


\bibitem{MOF1}%\cite{Mota:2003tc}
%\bibitem{Mota:2003tc}
  D.~F.~Mota and J.~D.~Barrow,
``Varying alpha in a more realistic Universe,''
  Phys.\ Lett.\ B {\bf 581}, 141 (2004)
  doi:10.1016/j.physletb.2003.12.016
  [astro-ph/0306047].
  %%CITATION = doi:10.1016/j.physletb.2003.12.016;%%
  %188 citations counted in INSPIRE as of 06 Dec 2018



   \bibitem{ch19} %\cite{Saaidi:2011zza}
%\bibitem{Saaidi:2011zza}
  K.~Saaidi, A.~Mohammadi and H.~Sheikhahmadi,
``$\gamma$ Parameter and Solar System constraint in Chameleon Brans Dick theory,''
  Phys.\ Rev.\ D {\bf 83}, 104019 (2011)
  doi:10.1103/PhysRevD.83.104019
  [arXiv:1201.0271 [gr-qc]].
  %%CITATION = doi:10.1103/PhysRevD.83.104019;%%
  %20 citations counted in INSPIRE as of 06 Dec 2018


    \bibitem{ch20} %\cite{Saaidi:2013yfa}
%\bibitem{Saaidi:2013yfa}
  K.~Saaidi, H.~Sheikhahmadi, T.~Golanbari and S.~W.~Rabiei,
 ``On the holographic dark energy in chameleon scalar-tensor cosmology,''
  Astrophys.\ Space Sci.\  {\bf 348}, 233 (2013)
  doi:10.1007/s10509-013-1491-5
  [arXiv:1404.2139 [gr-qc]].
  %%CITATION = doi:10.1007/s10509-013-1491-5;%%
  %5 citations counted in INSPIRE as of 06 Dec 2018


   \bibitem{ch1:20a}%\cite{Aghamohammadi:2013eja}
%\bibitem{Aghamohammadi:2013eja}
  A.~Aghamohammadi, K.~Saaidi, A.~Mohammadi, H.~Sheikhahmadi, T.~Golanbari and S.~W.~Rabiei,
``Effect of an external interaction mechanism in solving agegraphic dark energy problems,''
  Astrophys.\ Space Sci.\  {\bf 345}, no. 1, 17 (2013)
  doi:10.1007/s10509-013-1386-5
  [arXiv:1402.2608 [physics.gen-ph]].
  %%CITATION = doi:10.1007/s10509-013-1386-5;%%
  %6 citations counted in INSPIRE as of 06 Dec 2018



\bibitem{Planck2015} %\cite{Ade:2015lrj}
%\bibitem{Ade:2015lrj}
  P.~A.~R.~Ade {\it et al.} [Planck Collaboration],
``Planck 2015 results. XX. Constraints on inflation,''
  Astron.\ Astrophys.\  {\bf 594}, A20 (2016)
  doi:10.1051/0004-6361/201525898
  [arXiv:1502.02114 [astro-ph.CO]].
  %%CITATION = doi:10.1051/0004-6361/201525898;%%
  %1828 citations counted in INSPIRE as of 06 Dec 2018






\end{thebibliography}
\end{document}